\documentclass[12pt,preprint]{aastex}

\shorttitle{Characterizing the AO Off-Axis PSF. II}
\shortauthors{Steinbring et al.}

\input epsf
\def\plotone#1{\centering \leavevmode
\epsfxsize=1.0\columnwidth \epsfbox{#1}}

\def\plotonehalf#1{\centering \leavevmode
\epsfxsize=0.5\columnwidth \epsfbox{#1}}

\begin{document}

\title{Characterizing the Adaptive Optics Off-Axis Point-Spread Function. II. Methods for Use in Laser Guide Star
Observations\altaffilmark{1,2}}

\author{E. Steinbring\altaffilmark{3}, S.M. Faber\altaffilmark{4}, B.A. Macintosh\altaffilmark{5}, D.
Gavel\altaffilmark{4}, and 
E.L. Gates\altaffilmark{6}}

\altaffiltext{1}{Based on observations obtained at the Lick Observatory, which is operated by the University of
California.}
\altaffiltext{2}{All authors are affiliated with the Center for Adaptive Optics.}
\altaffiltext{3}{Herzberg Institute of Astrophysics, National Research Council Canada, Victoria, BC V9E 2E7, Canada}
\altaffiltext{4}{UCO/Lick Observatory, University of California, Santa Cruz, CA 95064}
\altaffiltext{5}{Lawrence Livermore National Laboratory, P.O. Box 808, Livermore, CA 94551}
\altaffiltext{6}{Lick Observatory, P.O. Box 85, Mt. Hamilton, CA 95140}
 
\begin{abstract}

Most current astronomical adaptive optics (AO) systems rely on the availability of
a bright star to measure the distortion of the incoming wavefront. Replacing the guide star
with an artificial laser beacon alleviates this dependency on bright stars and therefore 
increases sky coverage, but it does not eliminate another serious problem
for AO observations. This is the issue of PSF variation with time and field position near 
the guide star. In fact, because a 
natural guide star is still necessary for correction of the low-order phase error, characterization 
of laser guide star (LGS) AO PSF spatial variation is
more complicated than for a natural guide star alone. We discuss six methods
for characterizing LGS AO PSF variation that can potentially improve the determination of the PSF away 
from the laser spot, that is, off-axis.
Calibration images of dense star fields are used to determine the change in PSF variation with
field position. This is augmented by AO system telemetry and simple computer simulations to determine a
more accurate off-axis PSF. We report on tests of the methods using the laser AO system
on the Lick Observatory Shane Telescope.
We observed with offsets typical of separations between dim science targets and the nearest
suitably bright tip-tilt guide star, up to 20{\arcsec}. If the tip-tilt guide star is used as the PSF reference, the predicted Strehl ratio
within an 8{\arcsec} radius of the LGS can be more than a factor of two too high, due to not taking into
account the field dependent anisoplanatism.
A better result may be obtained by an improved empirical approach: repositioning the tip-tilt 
guide star to compensate for the additional 8{\arcsec} offset.  This would result in a
27{\%} relative error in Strehl ratio. A simple Gaussian model of PSF variation
can 
reduce the error in the prediction to $\sim$20{\%}.
Some further improvement may by obtained semianalytically, by using telemetry from the AO system (14{\%} error) plus a 
more sophisticated theoretical model of PSF variation (13{\%} error). It should be noted, however,
that all but the first method achieve comparable accuracy in predicting PSF full-width at half maximum. This is 
important because the last method, unlike the others, is not directly applicable to 10-meter apertures without
further computational complexity.

\end{abstract}


\section{Introduction}\label{introduction}

An introduction to the problem of characterizing the PSF of a natural guide star (NGS)
adaptive optics (AO) system is given in \citet{Steinbring2002} (hereafter referred to
as Paper I). In an NGS observation a bright star is used to guide the AO system while the
telescope is pointed at a scientific target which is typically 10{\arcsec} - 30{\arcsec}
away, and perhaps as distant as 1{\arcmin}. Because the performance of the AO system degrades 
for fainter guide stars - correction
is poorer as signal-to-noise (S/N) on the wavefront sensor (WFS) decreases - any
subsequent calibration observation must use a guide star of similar brightness. Even
so, the delivered correction depends directly on quickly changing seeing conditions
(measured by the Fried parameter $r_0$) and is therefore not likely to be the same
during the two observations. Further aggravating this, the spatial dependence of the
phase error is dependent on the distribution of turbulence strength along the line of
sight (given by the $C_n^2(h)$ profile, where $h$ is height), and thus PSF
anisoplanatism can be strong and change with time. Therefore a calibration measurement
done with the same offset but at a later time may provide little information about the
PSF at the scientific target. In Paper I we discussed a method which may improve the
determination of the AO off-axis PSF. It uses a mosaicked observation of a dense star
field to characterize the field dependence of the PSF. The entire mosaic is deconvolved
with the image of the guide star in order to generate a ``kernel map". Each deconvolved
star is a function which can convert the guide star image into a PSF at that off-axis
position. The kernel map is combined with on-axis PSF information during the target
observation in order to produce a PSF estimate that is potentially more accurate for both the
target position and time of the scientific observation.

The technique we outlined in Paper I provided modest but useful improvement in the accuracy of NGS AO off-axis
PSF reconstruction - reducing a 60{\%} error in the prediction of FWHM at 25{\arcsec} offset (if one used just the 
guide star image alone) down to about 19{\%}
by our method. For many scientific programs this level of accuracy would be helpful. For example, it would
assist in subtracting the pointlike core from the image of a quasar host or radio galaxy. The method would be
particularily helpful in
this case because one is very unlikely to find a star sufficiently bright to guide AO within a few arcseconds of
any given galaxy. Worse, even if a usable star is within 30{\arcsec}, it is likely that there is still no 
true ``PSF star", that is, a star 
within a few arcseconds of the target sufficiently bright to provide a simultaneous measurement of the target PSF.
The advent of laser beacons for AO promises to overcome the problem of guide star availability by
providing a bright artificial ``star" which can be positioned near any target. 
Having a pointable guide star does not, however, solve the problem of PSF characterization. In those cases where 
no PSF star is available - just as in the case of NGS AO - some other form of PSF characterization will be needed.

In laser guide star (LGS) AO, the artificial beacon is produced either by Rayleigh scattering or 
the excitation of atmospheric sodium at high altitude. The outgoing laser beam is
usually sighted along the optical axis of the telescope, which means that in a typical observation 
both are pointed at the scientific target. The ``laser spot" which it produces is entirely optical emission, and thus 
after reflecting off the deformable mirror (DM) a dichroic
is used to send this light to a WFS and allow the NIR light to pass through to the scientific camera.
These optics are essentially unchanged from the NGS AO case. 
However, the laser beam wanders by an unknown amount on the way up and thus the absolute position of the laser spot on
 the sky is unknown, and it can only be used to sense the
high-order aberrations of the incoming wavefront. An NGS is still needed to probe global tilts across the
pupil. This can be done by means of a separate quad-cell sensor for the NGS, for example. 
The radial dependence of anisoplanatism for low Zernike orders is weak, and thus the NGS 
and LGS need not be at the same position on the sky. In fact, it is highly desirable to
guide with the laser pointed at the scientific target - because it takes advantage of
the best correction - and choose a convenient nearby (less than $\sim1${\arcmin} away) star for the quad-cell.
In this paper we only consider data where the LGS is coincident with the telescope optical axis, and so we will 
consistently refer to the location of laser spot as on-axis.
One could still obtain a direct on-axis PSF measurement by interleaving scientific observations
with short exposures of the tip-tilt star with the laser (and telescope) pointed at it.
But if the offset is larger than a few arcseconds this method will provide a poor
estimate of the off-axis PSF, especially if anisoplanatism is strong, as is the case for $C_n^2$ profiles skewed to
high altitude. 
A better method would incorporate some means of modeling this anisoplanatism into a corrective off-axis PSF kernel.
The results could be limited by how quickly and often the tip-tilt guide star calibration exposures were taken,
however,
because the seeing could change significantly over the course of a scientific observation. 

The differences between the optical configurations of LGS and NGS AO suggest that the semiempirical method of Paper I 
for characterizing the PSF is not applicable for LGS observations, for two reasons. 
First, the field dependence of the PSF
variation will be different. Second, a typical LGS AO scientific observation will provide no simultaneous on-axis
measurement with which to calibrate the seeing variations of the PSF. The difference in
field dependence for LGS AO is easy to understand. It is partly due to the finite
altitude of the artificial beacon, which is perhaps only 15 km for one produced by
Rayleigh backscatter, or 90 km for emission from excited atmospheric sodium. That is,
the volume of turbulent air traversed by the return beam does not include the entire
atmosphere nor is it a cylinder - as it would be if the LGS were effectively at
infinity. This ``cone effect" degrades the on-axis LGS Strehl ratio, but it is not the most 
important difference in field dependence compared to NGS AO, and it is a weaker effect with smaller aperture.
A more important issue is the angular separation of the LGS and NGS on the sky. Even
though the NGS can be much fainter than that needed for sensing high-order aberrations,
in practice the nearest available star may be up to 1{\arcmin} away. Clearly, the larger the separation between the
LGS and NGS the more complicated the field dependence will be. The separation of the LGS
and NGS also causes a more serious problem with the application of our previous method.
In an LGS observation the on-axis PSF is now at the position of the scientific
target, which may not be a star. The laser spot is invisible in the NIR and therefore
provides no measurement of the on-axis PSF in the scientific image.

We describe six techniques for determining the LGS AO off-axis PSF and 
apply these methods to observations made with the Shane telescope AO system at Lick
Observatory.
This is a continuation of our program funded by the Center for Adaptive
Optics to study AO PSFs, which in Paper I dealt with NGS data only. The results of the
LGS methods are of accuracy similar to the semiempirical method of Paper I, $\sim$20{\%} accuracy in 
Strehl ratio for offsets up to 28{\arcsec}. In
Sections~\ref{strategy} and~\ref{map} we 
propose an observing strategy for LGS-mode PSF characterization and discuss the functional form of the kernel map. 
In Section~\ref{observations}
we describe the observations and data reduction. Analytical calculations and computer simulations were
used to further characterize the data, and the PSF prediction methods were applied to them. 
These analyses are found in
Section~\ref{analysis}, and our conclusions follow in Section~\ref{conclusions}.

\section{Observing Strategy}\label{strategy}

We have described how LGS and NGS AO observations differ, but at least their high-order corrections should be similar.
In Paper I we noted that for perfect correction of $N$ Zernike modes the Strehl ratio at offset $\theta$ from the
guide star should be given by $$S=S_0\exp[-(\theta/\theta_N)^{5/3}], \eqno(1)$$ with the
exponent tending toward two for very partial correction. The ``instrumental"
isoplanatic angle $\theta_N$ depends on the Fried parameter $r_0$ and mean height of
turbulence $\bar{h}$ as $$\theta_N=0.314{{r_0 \cos{\gamma}}\over{\bar{h}}}, \eqno(2)$$
where $\gamma$ is the zenith angle. If we neglect tip-tilt anisoplanatism, equations
1 and 2 should still describe the high-order LGS AO correction (with $\theta$ now refering to radius
from the LGS) because cone effect is folded into the value of $S_0$.  
In reality, wavefront tip and tilt are sensed differently in LGS compared to NGS mode, so the analogy is not exact. 

If we assume cone effect and tip-tilt anisoplanatism are negligible, a typical
LGS AO observation then differs from an NGS one only by the absence of an on-axis PSF star.
We can we still determine the on-axis Strehl ratio.
A simple model of the LGS AO on-axis Strehl ratio might have it 
degraded by a wavefront fitting error, a delay error, and photon statistics. This
is $$S_0=\exp{(\sum_{\rm all~sources}{-\sigma^2})}=\exp{[-(n_0/n)^{5/6}-(\tau/\tau_0)^{5/3}-p_0/p]},\eqno(3)$$
assuming the errors are uncorrelated, where $n$ is the number of actuators, $\tau$ is the delay time, $p$ is the guide
source
flux, and the following definitions are used (see, for e.g., Roddier 1999):
$$n_0=0.27(D/r_0)^2,\eqno(4)$$ $$\tau_0=0.314r_0/\bar{v}, \eqno(5)$$ and
$$p_0=2.67\times 10^{(3-0.4V)/7}\tau q\lambda^2/r_0^2 \eqno(6)$$ for telescope diameter
$D$, mean windspeed $\bar{v}$, guide-star magnitude $V$, throughput $q$, and
wavelength $\lambda$. Here $q$ is the total throughput of the telescope plus AO system.
Note that even this ``simple" model requires several atmospheric
measurements and AO system telemetry to estimate $S_0$.
We can still try to use it, but it makes more sense to find $S_0$ by fitting equation 1 to the 
star-field data, as we did in Paper I. This will account for cone effect by interpolating an estimate of $S_0$. 

Fitting equation 1 to the star-field data will also provide an estimate of $\theta_N$.
To characterize $\theta_2$ we must fit equation 1 to a further 
set of star-field mosaics - this time with high-order correction turned off.

A proposed observing strategy for the LGS AO PSF characterization is similar to 
that described in Paper I, with one new step. The observer obtains images of a dense star field once or
at most a few times per night; the core of a globular cluster would work best. The AO system
is guided on a suitable tip-tilt guide star. The goal is
to obtain unsaturated images of a roughly uniform distribution of stars from which to determine the 
isoplanatic angle. To keep the analysis straightforward, the same offsets between the LGS and tip-tilt guide star used
in the
scientific observations should be repeated here. The contribution of tip-tilt anisoplanatism to the
PSF is measured in a separate mosaic by repeating the mosaic observations with the high-order
 correction turned off. Obtaining this tip-tilt-only mosaic is an extra step compared to Paper I. 
The tip-tilt-only and high-order mosaics are used together to 
determine the PSF variation by fitting each independently with equation 1. The off-axis PSF
during a scientific observation is provided by applying an appropriate kernel to either the observed PSF at the
position of 
the tip-tilt guide star or one generated using WFS telemetry.

\section{Models of the Kernel Map}\label{map}

Unlike in Paper I, to obtain a two-dimensional model of the LGS AO PSF, we must theorize
the functional form of the kernel map. For simplicity, let us assume for now that tip-tilt 
anisoplanatism is negligible.
The off-axis PSF then has the 
form $${\rm PSF}_{\theta}({\mathbf x})={\rm
PSF}_0({\mathbf x})\ast{\rm K}_{\theta}({\mathbf x}) \eqno(7)$$ where ${\mathbf x}$ is a position
in the focal plane, K is the kernel, and $\ast$ indicates a convolution. Note that since ${\rm
PSF}_0$ is the on-axis PSF we need not assume that cone effect is negligible. The optical
tranfer function (OTF) is the Fourier transform of the PSF and will have associated
with it an ``anisoplanatic transfer function" (ATF), the Fourier transform of the
kernel. Equation 7 in terms of spatial frequencies ${\mathbf f}$ is then $${\rm
OTF}_{\theta}({\mathbf f})={\rm OTF}_0({\mathbf f})\cdot{\rm ATF}_{\theta}({\mathbf
f}). \eqno(8)$$ Our goal is to find the functional form of either K or ATF.

\subsection{Gaussian Kernel}\label{gaussian_kernel}

The results of Paper I suggest that the NGS AO kernel is approximately Gaussian with a FWHM varying 
with $\theta$. Let us
assume that the LGS AO kernel also has this shape. That is, $${\rm
K}_{\theta}(r)=S\exp{[-(r/{\rm FWHM})^2]}. \eqno(9)$$ The volume of the kernel must be
unity everywhere, which gives $$1=S\int_0^{2\pi}{\rm d}\phi\int_0^{\infty}{\rm
d}rr\exp{[-(r/{\rm FWHM})^2]}. \eqno(10)$$ Thus if we assume that the kernel has a
width of one pixel on-axis we have $${S\over{S_0}}={\big(}{1\over{\rm FWHM}}{\big)}^2, \eqno(11)$$ where FWHM is in
pixels. This ``Gaussian kernel" is obviously a
simplification of the kernel shape because a Gaussian of one pixel in width is not a true delta function. Let us 
just assume for now that this is a reasonable approximation. We have not assumed that cone effect is
negligible. For any non-zero offset $\theta$, equation 1 combined with equations 9 and 11 then suggests that 
the kernel should be $${\rm
K}_{\theta}(r)=\exp{[-r^2\exp{[-(\theta/\theta_N)^{5\over{3}}]}]}. \eqno(12)$$

\subsection{Theoretical Kernel}\label{theoretical_kernel}

It is much more mathematically complex, but one can also determine the kernel from theory. We can decompose the
turbulent phase onto a Zernike polynomial basis and see how this decomposition changes with
increasing pupil offset. For a given offset the phase can be written as
$$\Phi_{\theta}({\mathbf r})=\sum_{i=2}^{\infty}a_i(\theta)Z_i({\mathbf r}) \eqno(13)$$
where $a_i(\theta)$ is the Zernike coefficient at offset $\theta$. 
We will assume that the AO system perfectly corrects $N$ Zernike modes. The residual
phase is then $$\Phi_{{\rm res}, {\theta}}({\mathbf
r})=\sum_{i=2}^{N}(a_i(\theta)-a_i(0))Z_i({\mathbf
r})+\sum_{i=N+1}^{\infty}a_i(\theta)Z_i({\mathbf r}). \eqno(14)$$ 
\citet{Fusco2000} have shown in this circumstance that the ATF can be written as $${\rm
ATF}_{\theta}({\mathbf f})=\exp{[-{1\over{2}}{\rm SF}_{\theta}(\lambda{\mathbf f})]}, \eqno(15)$$ where SF is the
residual phase structure function
for wavelength $\lambda$
given by $${\rm SF}_{\theta}(\lambda{\mathbf
f})=\sum_{i=1}^N\sum_{j=1}^{\infty}2[<a_i(0)a_j(0)>-<a_i(0)a_j(\theta)>]U_{i,j}(\lambda{\mathbf
f}), \eqno(16)$$ and the $U_{i,j}$ are functions defined as $$U_{i,j}(\lambda{\mathbf
f})={\int {\rm d}{\mathbf r}[Z_i({\mathbf r})-Z_i({\mathbf r}+\lambda{\mathbf f})][Z_j({\mathbf
r})-Z_j({\mathbf r}+\lambda{\mathbf f})]P({\mathbf r})P({\mathbf r}+\lambda{\mathbf
f})\over{\int {\rm d}{\mathbf r}P({\mathbf r})P({\mathbf r}+\lambda{\mathbf f})}}, \eqno(17)$$ 
for position ${\mathbf r}$ over the range of pupil function $P$. The angular correlations $<a_i(0)a_j(\theta)>$ 
of this ``theoretical kernel" have been
computed by \cite{Chassat1989} and \cite{Molodij1998} by assuming particular $C_n^2$ profiles, and 
implicitly assuming that cone effect is negligible. These correlations are
$$<a_i(0)a_j(\theta)>=3.895\Big{(}{D\over{r_0}}\Big{)}^{5\over{3}}{{\int_0^{h_{\rm atm}}{\rm
d}hC_n^2(h)I_{l,m}({{2\theta
h}\over{D}})\over{\int_0^{h_{\rm atm}}{\rm d}hC_n^2(h)}}}, \eqno(18)$$ where $h_{\rm atm}$ is a height sufficient to
be beyond atmospheric
turbulence. The integrals $I_{l,m}$ can be calculated either by numerical means or by the method of Mellin transforms; the 
latter can be found in \cite{Molodij1998}. Here $l$ and $m$ correspond to the radial
orders of Zernike modes $i$ and
$j$.

Although the Gaussian kernel seems to be applicable to both NGS and LGS modes, some error will be committed by
neglecting cone effect 
in the theoretical kernel. Even so, for 3-meter aperatures it is a small error. \citet{Molodij1997} have calculated
the
residual phase variance for perfect correction of all Zernike modes ($N=\infty$) over a circular
aperture when the incoming wavefronts are spherical, that is, propagating from a height of 15 or 90 km. 
For a Hufnagel profile with a mean turbulence height of $\bar{h}=6.7$ km and an LGS at 90 km they find
a residual phase variance of $$\sigma^2=0.006(D/r_0)^{5/3}. \eqno(19)$$ If one applies
the Marechal approximation this suggests perfect LGS AO correction on a 3 meter could at best achieve an on-axis
Strehl ratio of 
$1/\exp(0.0374)=0.96$ at $K$, which is very close to unity. Put another way, everything else being equal, this system
should achieve about 4\% less
Strehl ratio
than an NGS system. A loss of Strehl ratio of 4\% due to the cone effect is not a serious concern
for our data, where Strehl ratios are more typically 30\%. This would not be true for AO on a 10 meter, however,
because 
a similar calculation suggests that the cone effect results in a degradation
of about 24\% in Strehl ratio. Thus the theoretical kernel may not be directly applicable to Keck 
AO without first accounting for the cone effect. This is important because the integrals in equation 18 
require significant computational effort. We will show that for our data this complex analytical kernel is not
needed, and 
the simpler Gaussian kernel - which accounts for cone effect - gives comparable results.

\subsection{Tip-Tilt Kernels}

So far we have modeled our kernel maps by assuming that tip-tilt anisoplanatism is negligible.
A less severe and more plausible assumption is that the tip-tilt and high-order loops are independent. 
This will allow us to develop two auxiliary kernels with which we may conceivably account for tip-tilt anisoplanatism.
Equation 12 should apply equally well to NGS mode, so the Gaussian ``tip-tilt Gaussian kernel" has an exponent of two,
according to equation 1.
Also, equation 18 is actually a calculation for NGS mode. So it 
follows that the ``tip-tilt theoretical kernel" would be given by equation 16 with $N=2$.
Note that $\theta$ now refers to the offset from the NGS.
To account for the tip-tilt component of the field variation we should be able to degrade the tip-tilt guide-star PSF 
with either of these modified kernels.
We could next apply the high-order LGS Gaussian or theoretical kernels to this map as a second step, and see it this
reproduces 
the observed LGS field variation. 

\section{Observations and Data Reduction}\label{observations}

Observations were made with the 3.0 m Shane Telescope of the Lick Observatory using the
Lawrence Livermore National Laboratory (LLNL) AO system in LGS mode. We discussed the
operation of the AO system in NGS mode in Paper I.
See \citet{Bauman1999} and \cite{Gavel2000} for further discussion of the NGS system; 
\citet{Olivier1999} and \cite{Bauman2002}
for a description of the LGS configuration. The artificial beacon is produced by a
pulsed solid-state pumped-dye laser tuned to excite the 589 nm resonance line of
atomic sodium, which is naturally present in an atmospheric layer at about 90 km in
altitude. The laser typically operates at about 10 watts output, although up to 15 watts
has been achieved. The return power depends on seasonal variation of
density and thickness of the sodium layer, nightly variation of intervening thin cloud, and 
target airmass.  The laser launch telescope is attached to the side of the 
Shane and produces a spot along the optical axis, aligned with the same CCD-based Shack Hartmann sensor used
in the NGS system. During our observations the LGS was sufficiently bright to permit
CCD framerates of 100 Hz, about twice as fast as during our NGS run. We always used a
relatively bright ($V=12$) star for global tip and tilt measurement ($V<16$ is
possible). The tip-tilt quad cell can patrol within a region roughly
2{\arcmin}-square, which permits telescope offsets of up to 55{\arcsec} in any
direction. The quad cell drives a tip-tilt mirror; higher order aberrations are
corrected with the same 61 active element piezostack mirror used in NGS mode. 

\begin{figure}
\plotone{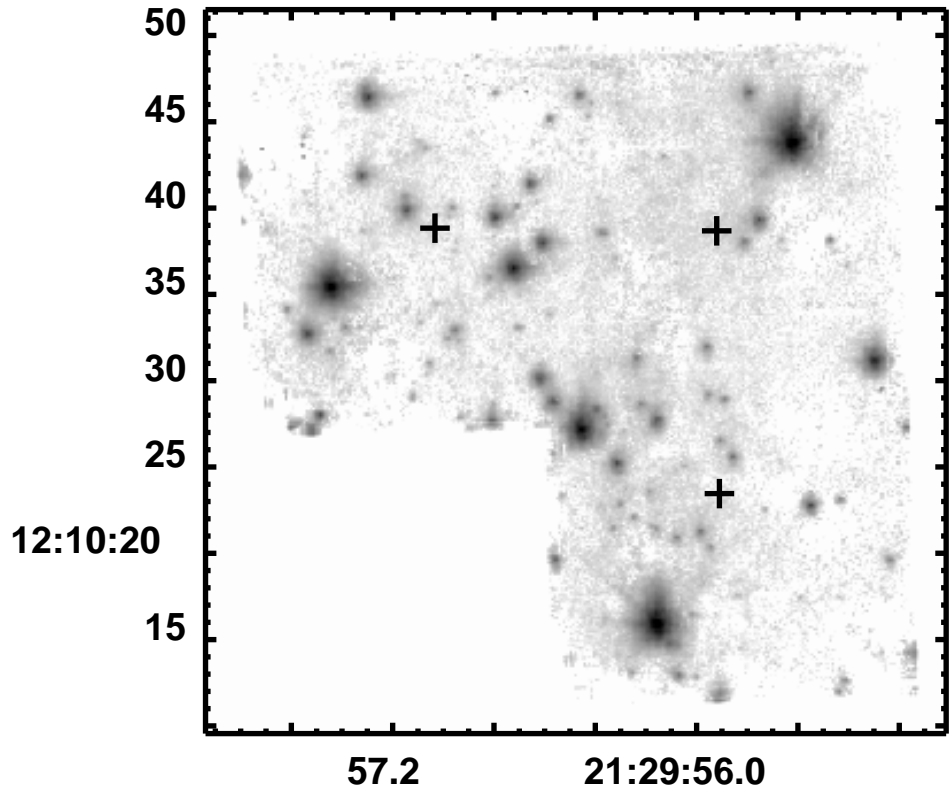}
\caption{A single mosaiced observation of M15 from the Lick LGS AO dataset. North is up
and east left in this log grey-scale image. Right ascension and declination are given
in J2000.0 coordinates. The mosaic is composed of three pointings, with some overlap. A
main pointing is flanked by two others, one to the south, and one to the east. The
center of each pointing is coincident with the laser spot, which is indicated by a
cross. The tip-tilt guide star is the bright star at upper right.}
\label{figure_m15}
\end{figure}

Our LGS observations of M15 were obtained through a $K_{\rm s}$ filter
with the Infrared Camera for Adaptive Optics at Lick (IRCAL) \citep{Lloyd2000} in
August 2001. The same tip-tilt guide star was used for all observations - the same
source used for the NGS observations of this field in September 2000 - which had the advantage of
being in a region of sky already mapped by us. Archival $V$, $R$, and $I$-band Hubble
Space Telescope (HST) Wide-Field Planetary Camera 2 (WFPC2) images were also available.
The large 2.6{\arcmin} $\times$ 2.6{\arcmin} FOV of WFPC2 helped to provide astrometry
for registration of the mosaicked AO images. Our previous NGS observations had shown
that the core of the cluster, 45{\arcsec} to the southeast, was too crowded to obtain
reliable Strehl ratio measurements for AO. This region was therefore avoided. Three individual
pointings of the telescope produced a chevron shaped mosaic roughly 35{\arcsec} across 
along each arm. Figure~\ref{figure_m15} is an image of the resulting field. The crosses
indicate the position of the laser spot in each pointing, and the tip-tilt guide star is the
bright star at upper right. Notice the triangular arrangement of bright stars in this
field.
One can see that we could have obtained a direct measurement of the on-axis PSF in every frame by centering
each observation on a vertex of this constellation. 
To allow a test that will be explained 
in Section~\ref{generating_artificial_images}, we instead opted to offset the laser either 8{\arcsec} or 
20{\arcsec} from the tip-tilt star so as to maintain the same 8{\arcsec} separation
between the laser pointing and each PSF star.
Reducing the largest offset from 25{\arcsec} to 20{\arcsec} also made our
observing sequence faster. The exposure time at each
pointing was 100 seconds, and each mosaic took only approximately 10 minutes to complete.
Every third or fourth pass was observed with the high-order correction turned off,
that is, it was obtained with tip-tilt only correction. A total of 25 mosaics,
including the tip-tilt only fields, was obtained over two nights. A journal of the
observations is given in Table~\ref{table_journal_of_observations}.

No independent measurements of seeing or the turbulence profile were available at Lick.
We were still able, however, to determine $r_0$ during the run.
Wavefront-sensor telemetry was recorded at intervals by the AO system and used to
determine $r_0(0.5 \mu{\rm m}, {\rm zenith})$ in a manner similar to \citet{Veran1997}.
Each mosaic observation was bracketed by $r_0$ measurements, and the averages of those
are given in Table~\ref{table_journal_of_observations}. Note that the seeing during the first 
night ($\bar{r}_0=20$ cm) was almost a factor of two better than the second night ($\bar{r}_0=12$ cm). 

The output laser power at the beginning of the
first night was 13.4 watts. No output power was recorded for the second night, but it was probably
similar. Variation in the returned signal over the course of each night can be inferred from the 
mean WFS illumination. The standard deviation of the 
mean return power was 7\% for the first night and 35\% for the second.  Some unknown part 
of this variation could have come from changes in output power.  Variable thin cloud could also 
have played a role, but apparently a small one. The flux from the tip-tilt guide star was measured in each of the
mosaics and was stable over both nights to within 4\% (std. dev.). All observations were at an 
airmass less
than about 1.3, which leaves only fluctuations in the concentration of atmospheric sodium as a 
remaining possible source
of LGS variability.  Perhaps this accounts for
the larger changes in return power on the second night. 

The windspeed was recorded by an anemometer on the roof of a nearby telescope building, situated about 
100 m from
the Shane dome. The anemometer measurements may be affected by the sheltering of buildings and terrain, and
thus the resulting average $\bar{v}=1.8~{\rm m}~{\rm s}^{-1}$ almost certainly underestimates the 
true windspeed of the bulk of turbulence. An atmospheric
profiling instrument such as a Scintillation Detection and
Ranging (SCIDAR) device, if it had been available, could have provided an independent
measurement of $\bar{h}$ and estimates of the windspeed at higher altitudes. 

All the globular-cluster data were compared to HST WFPC2 observations of the same
fields in order to determine the plate scale and orientation of the
cameras. We obtained good relative astrometry of stars, with sub-pixel accuracy. The
data were flat-fielded, and aperture photometry was carried out on all the
sky-subtracted images. 

Strehl ratios were calculated in the same manner as described in Paper I. The accuracy of
these values is similar to the photometric accuracy, which was better than 4{\%}. We obtained an
estimate of the on-axis Strehl ratio $S_0$ and the instrumental isoplanatic angle
$\theta_N$ by a least-squares fit of equation 1 to the measurements of Strehl ratios
for stars in each frame of the mosaics. For the tip-tilt only correction frames, the 
fitted values of $S_0$ agreed well with the direct
measurement of the guide star. This direct measurement was not available in the the LGS observations.
The values of both $S_0$ and $\theta_N$ corrected for
zenith angle are given in Table~\ref{table_journal_of_observations}.

It is somewhat surprising that $S_0$ in most of the 20{\arcsec} offset fields are higher than in 8{\arcsec} offset fields taken at 
almost the same time, although this is not always the case. One would expect exactly the opposite due to tip-tilt anisoplanatism. 
There are fewer stars in the 8{\arcsec} pointing, which does contribute to a noisier fit with equation 1. There
are also correspondingly fewer stars very close to on-axis, which might lead to a small systematic underestimate of $S_0$, because the fit will be
poorest at the most sharply falling part of the curve.

The instrumental isoplanatic angles (at 2.2 $\mu{\rm m}$) were approximately
13{\arcsec} for LGS and 32{\arcsec} for tip-tilt on average. We alternated between tip-tilt only and full correction
mosaics during the observations and thus both should sample roughly the same $\bar{h}$. This suggests that 
$\theta_2\approx2.5\theta_N$. Combining these values of
$\theta_N$ with the mean value of $r_0$ yields an estimate of $\bar{h}=5$ km via equation 2.

The shape of the mosaics should allow us to detect azimuthal variation in anisoplanicity, if it were
present, by comparing $S_0$ in the eastern and southern arms. This second-order anisoplanicity might be expected if
the
windspeed $\bar{v}$ were sufficiently fast to outpace the bandwidth of the AO system more in one direction than
another.
We do not detect a strong azimuthal anisoplanatism here. The average relative difference in $S_0$ between eastern
and southern arms is only about
7{\%} (an absolute difference in Strehl ratio of $\Delta S=1${\%}) for the first
night, which is well within the observational uncertainties. This difference increased to about 19{\%} ($\Delta
S=3${\%}) for the second night,
again not significantly larger than the observational uncertainty.

\section{Analysis}\label{analysis}

Let us first compare the data to the expected performance given by equations 1 and 3.
We will assume $N=9$, $2~{\rm km}<\bar{h}<8{\rm km}$, $1.00<q<0.33$, $\tau\approx(1/{\rm framerate})=0.01$ s, and that
$\bar{v}$
was either the mean value that we recorded on the ground or a factor of 2 faster. To simplify the calculations we
approximated the 
laser beacon by a star of magnitude $V=9$. An A-type star of this brightness provides roughly the
same number of photons per subaperture as the laser at its typical output of 10 watts.
Note that $q$ is the total throughput of the entire AO system, and thus here it folds in the efficiency of the LGS
return. A range of $q$
should therefore give an optimistic and pessimistic estimate of $S_0$ because we did not record the true brightness of
the the laser spot for
every frame.

\begin{figure}
\plotonehalf{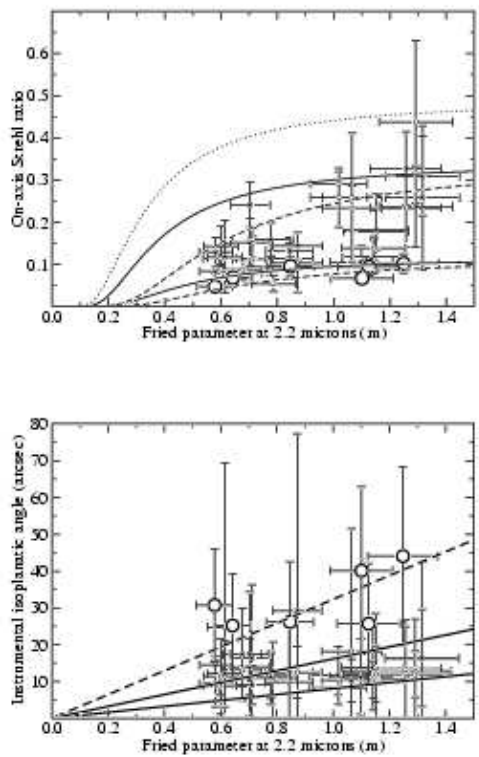}
\caption{Plots of on-axis Strehl ratio (top) and instrumental isoplanatic angle (bottom) as a
function of Fried parameter (filled circles). The tip-tilt only corrected data are indicated by open circles.
 Overplotted are the theoretical predictions. For
Strehl ratio, the dotted line
indicates a model with $q=1$, and solid lines the results for $q=0.67$ (upper) and $q=0.33$ (lower). The dashed lines
enclose the 
same range of $q$ as the solid lines but for a windspeed twice that measured on the ground (3.6 m ${\rm s}^{-1}$ vs
1.8 m ${\rm s}^{-1}$). For
isoplanatic angle, the upper solid line
assumes $\bar{h}=4$ km, and the lower solid line $\bar{h}=8$ km. The dashed line is for $\bar{h}=2$ km.}
\label{figure_theory}
\end{figure}

The LGS AO on-axis Strehl ratio data as a function of $r_0$ are shown in the upper panel of Figure~\ref{figure_theory}
as 
filled circles. For comparison, the tip-tilt only data are indicated by open circles. 
The model prediction assuming $q=1$ is overplotted as a dotted line. To achieve a Strehl ratio at or above
this line would require either
perfect throughput from the AO system or that the LGS was brighter than 10 watts. Therefore the one noisy data point
at $r_0\approx1.3$ m may be an
outlier. All of the remaining data are reasonably well bounded by the optimistic, $q=0.67$ estimate (upper solid line)
and
pessimistic, $q=0.33$ estimate (lower solid line) of $S_0$ which the model provides. We cannot distinguish between
a model with $v=1.8~{\rm m}{\rm s}^{-1}$ (solid lines) and one with $v=3.6~{\rm m}{\rm s}^{-1}$ (dashed lines). 

The lower panel of Figure~\ref{figure_theory} shows the measurements of instrumental isoplanatic angle. 
The symbols are the same as the upper panel. The LGS data appear to be well bounded by models with $\bar{h}=8$ km line
(lower solid line)
and $\bar{h}=4$ km line (upper solid line). The flatness of the $\theta_N$ distribution may be caused by 
a difference in $\bar{h}$ between the nights. This seems plausible because there was a sharp division at $r_0(2.2~\mu{\rm m})=0.9$ m
separating the first and second nights. The dashed line indicates a model with $\bar{h}=2$ km, which is within the
observational uncertainties
of most of the tip-tilt data. 

We confirm that the instrumental isoplanatic angle for tip-tilt
correction was large, about a factor of 2 or 3 larger than for full LGS operation. It is unclear, however,
how important the effect of partial tip-tilt correction is for the LGS results because
our simple model implicitly assumes perfect tip-tilt correction. It may be that we also cannot ignore the
effect that the finite angular separation of LGS and tip-tilt guide star has on the
instrumental isoplanatic angle. Only a more sophisticated model - one which includes imperfect tip-tilt correction - 
can determine the importance of this effect.  A description of one such model follows.
 
\subsection{Computer Modeling}\label{computer_modeling}

A computer simulation was developed to characterize the data. It is similar in principle to the simulation
described in Paper I, but necessarily somewhat more
complex. Here we require the simulation of two AO systems: the high-order AO loop, and the tip-tilt loop.
We used the Code for Adaptive Optics Systems (CAOS) software package \citep{Fini2001, Carbillet2001}, which is
available from the Observatorio
Astrofisico di Arcetri website. Like our previous simulation, it is written in IDL. We chose CAOS because 
it provides an intuitive graphical interface with which a realistic virtual AO system can be assembled quickly. The
user constructs the desired
optical configuration - including the appropriate offset between the NGS and LGS, for example - by arranging prebuilt
modules in
the GUI. System parameters such as detector pixel-scale and efficiency, framerate, and dwell time can be set. An
artificial
atmosphere is generated in a manner similar to that described in Paper I. The user supplies the desired number of
turbulent
layers, their heights, windspeeds and fractional contribution to $r_0$. The software then automatically determines the
appropriate system interaction matrices and constructs the simulation code. 

We assembled the simplest possible virtual AO system which still incorporated all of the essential components of the
real one. 
That is, we simulated a system with a 10 Watt sodium laser
beacon, a square Shack-Hartmann WFS, a 61 active-element piezostack deformable mirror, and a quad cell driving a
tip-tilt mirror. We assumed that
the $S/N$ of the high-order WFS was degraded only by Poisson noise and a readnoise of 10 ${\rm e}^-$ rms. The total
system throughput (telescope $+$ AO optics $+$ WFS detector) could be adjusted between 0.33 and 0.67.
We assumed that the quad-cell was noiseless.
The simulated DM and tip-tilt mirror were also very simple. We specified a DM that could perfectly reproduce the
shapes of up to the first 14
Zernike polynomials. We operated the DM without time delay, which
means that
all of the high-order wavefront error was caused purely by WFS noise and fitting error.
The tip-tilt mirror had a framerate limited to 100 Hz, which
means that all of the low-order wavefront error was due exclusively to finite bandwidth. 

The resulting AO
system is almost identical to the one which was successful in describing our NGS-mode data in Paper I, 
with the exception that tip-tilt and higher-order
correction have been
separated. This separation allows us to adjust the total throughput of the WFS and independently control the 
bandwidth of the tip-tilt loop.
The simplest possible atmospheric model was used. We allowed a single turbulence layer at $\bar{h}=3$, 5, or 7 km with
$r_0(0.5~\mu{\rm m})=12$ or 20 cm moving with a speed of 1.82 m ${\rm s}^{-1}$. The values of $r_0$ chosen were the observed mean
values for each night.
The goal is to see if a single-layer atmospheric model is a reasonable match to the LGS data, despite
the addition of a tip-tilt offset.

Here we have treated servo lag in the DM like a fitting error, which is admittedly a very simple treatment of the
high-order AO loop.
The reason for this is that the theoretical kernel treats anisoplanatism in the same way - as perfect correction of
$N$ Zernike modes. If we
can achieve a good fit with the simulation this should provide guidance for a value of $N$ in equation 16.

Because it was not included in our version of CAOS, we developed a new module which provides the integrated output PSF
at any
point in the field. This is essentially the output of the scientific camera. However, we did not take into account any
additional
aberrations due to this camera. Therefore, the outputs from our code should be the average values of the PSF Strehl
ratio
for the AO system alone, assuming perfect camera optics.

We ran simulations with $N=5$, 9, and 14.
For fixed $N$, raising or lowering $\bar{h}$ has the expected result of decreasing or increasing $\theta_N$.
The results for $N=9$ and $\bar{h}=5$ km are presented in Table~\ref{table_computer_modeling}.
The averages of the 8 and 20 arcsec offset models are in good agreement with the data, especially for the $r_0=12$ cm
model and the second night.
Models with $N<5$ are not
favoured because to obtain $\theta_N<16${\arcsec} with $N=5$ would require $h$ greater than 7 km.
Similarily, models with $N=14$ produce isoplanatic angles that are too small, even for $\bar{h}=3$. 

We would expect to obtain the best correction when the tip-tilt star and the LGS are
coincident. The models confirm this, and reproduce values similar to those predicted with equation 3. Recall that
equation 3
only applies for observations without offset, but we did not obtain any data for this configuration.
Even so, we can see that models without offset also plausibly fit the
data of the first night, although adding an offset brings $S_0$ into better agreement.  This is also true of the second night, although 
there the no-offset models predict isoplanatic angles which are too small. This suggests that although tip-tilt anisoplanatism is a weak
effect in our data it is not negligible.

In summary, the anisoplanatism predicted with the computer model is a reasonably good match to the data. 
The best fit models have $N\approx9$ and $\bar{h}=5$ km.
This does not preclude
variation in $h$ from night to night, or even within a night, but does suggest that a single layer of turbulence is a plausible model 
for both nights.
This model predicts a mean no-offset on-axis Strehl ratio of about 0.3 ($q=0.5$) over both of our nights. 
Note that if we had observed in this configuration, our on-axis Strehl ratio at
a tip-tilt offset of between 8{\arcsec} and 20{\arcsec} was just 0.18. This illustrates the
necessity for good PSF characterization for LGS
AO observations. One cannot rely on even a fairly nearby star to determine the local PSF. According to our model if
one were to use an
observation of the tip-tilt star (by aligning the LGS with it) to predict the on-axis PSF at at tip-tilt to LGS offset
of 20{\arcsec}, one would commit an error of at least 67{\%} in Strehl ratio. Using it to predict the PSF within a
radius of 8{\arcsec} from this position would increase the error beyond a factor of two. 

\subsection{Generating Artificial Images}\label{generating_artificial_images}

We now discuss three basic types of off-axis PSF predictions: empirical, semiempirical, and semianalytical. 
Variations within these catagories provide eight distinct methods.
We have just discussed the simplest empirical method, using the image of the tip-tilt guide star, which gives poor results.
One semiempirical
method was dealt with in Paper I and applies only to NGS data. This leaves six
LGS methods to be tested. Each carries with it assumptions about the importance of cone effect and tip-tilt
anisoplanatism.
Table~\ref{table_model_summary} gives a summary.

For each of the methods we generated artificial images for every LGS mosaic pointing. We did the same to 
the Lick NGS data discussed in Paper I, for those methods which are applicable. To 
provide a closer match to the LGS data we limited ourselves to those NGS data with on-axis Strehl ratio greater than
0.05. 
For comparison, we repeated the technique of Paper I on these better NGS data.

Images were generated as follows. For each pointing a blank 2-dimensional
array with the same dimensions as the original field was created. For each star in the field, an artificial PSF
with the same flux was generated. These were then inserted into the blank array at the same position as the original
star.

\subsubsection{Empirical Methods}

The propitious configuration of our LGS mosaics -
each laser pointing 8{\arcsec} from the vertex of a trianglular constellation of bright stars, 20{\arcsec} long on
each side - provided a
means of predicting the PSF at two locations in the field. We simply used the image of the
tip-tilt guide star in the first pointing to predict the PSF at the location of the bright star in the other two
pointings. This is plausible
because all three
of the bright stars are the same radius from their respective laser pointings. Thus one can think of this first-order
approach as having applied a
``corrective offset" to account for the high-order anisoplanatism. It assumes, however, that tip-tilt anisoplanatism
is negligible. 

\subsubsection{Semiempirical Methods}

\begin{figure}
\plotone{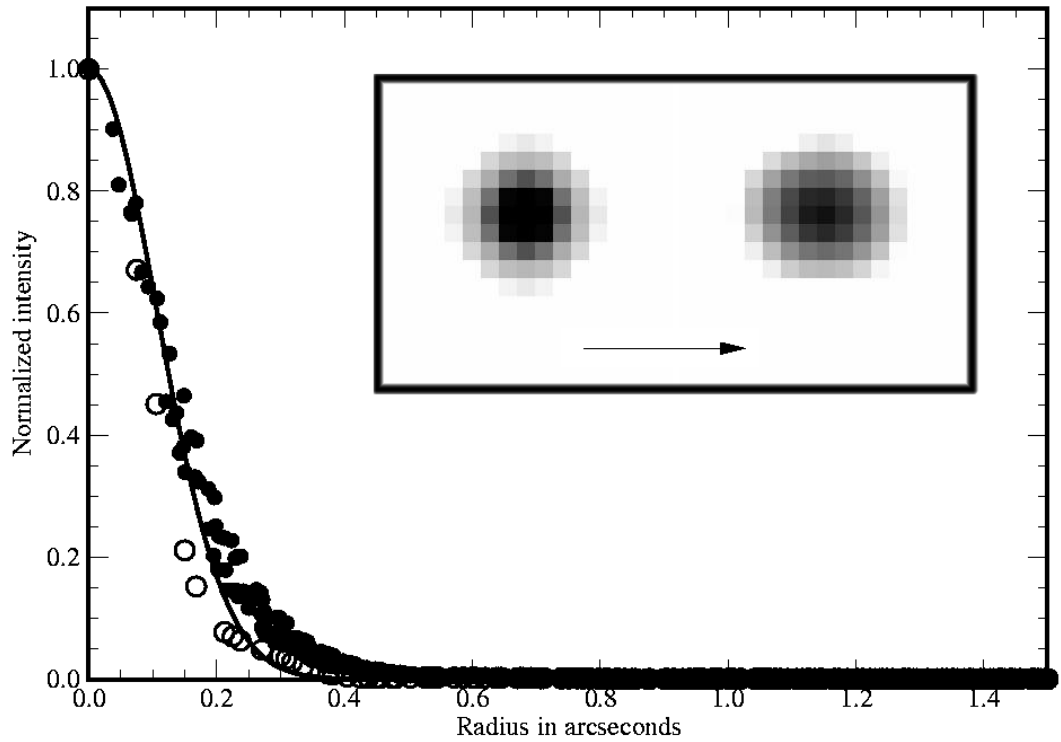}
\caption{Radial profiles of the tip-tilt kernels at 20{\arcsec} offset. The open circles indicate the Gaussian
kernel and the
theoretical kernel is given by the filled circles. Overplotted is a Gaussian curve fit to the theoretical model. This
is meant to show how similar
the two models are, especially in the core.  Inset are two dimensional
representations of
the Gaussian kernel (left) and theoretical kernel (right). The images have a FOV of 1{\arcsec} $\times$ 1{\arcsec} and
are displayed with a log
gray
scale. The direction to
the tip-tilt guide star is indicated by an arrow.}
\label{figure_kernel_tiptilt_20}
\end{figure}

To improve on the corrective offset method we added the second order effect of tip-tilt anisoplanatism by applying a
kernel (either the Gaussian or
theoretical one) to the tip-tilt guide-star image. These are the ``tip-tilt kernel" methods.
Figure~\ref{figure_kernel_tiptilt_20} shows the Gaussian and theoretical tip-tilt kernels for
offsets of 20{\arcsec}. The two kernels represent their respective averages of all the tip-tilt mosaic pointings.
Notice how similar they are,
especially in the core. The observed instrumental isoplanatic angle for
each field was used to determine the appropriate tip-tilt isoplanatic
angle, by applying the $\theta_2=2.5\theta_N$ correction mentioned in Section~\ref{observations}.
From this we can infer $\bar{h}$ from equation 2, which provides a single-layer model of $C_n^2$ for equation 18. 

Next, we tested the effect of just the high-order aberrations. This is called the ``high-order Gaussian kernel"
method. 
Here we have attempted to account
for tip-tilt anisoplanatism by predicting the PSF
based on the interpolated Strehl ratio at the center of the frame, that is, tip-tilt anisoplanatism should already be
included in the prediction of
on-axis Strehl ratio, and we need only correct for higher-order anisoplanatism.
To do this, we first took the interpolated values from equation 1 for
each of the off-axis pointings to find the peak Strehl ratio at the center of the frame. We then constructed a PSF
appropriate for the location of
the laser spot, because no star was available at this location. This PSF was estimated by 
convolving the telescope diffraction pattern with a Gaussian kernel assuming $S=S_0$ (the AO-corrected core), and then
adding the remaining flux
($1-S_0$) in a Gaussian of width $r_0/D$ (the seeing-limited halo). We then degraded this on-axis PSF for off-axis
positions by
further convolving it with the Gaussian kernel, using the observed instrumental isoplanatic angle. Note that the
on-axis PSF should 
already incorporate the degradation due to tip-tilt anisoplanatism at the location of the laser spot. 

\begin{figure}
\plotone{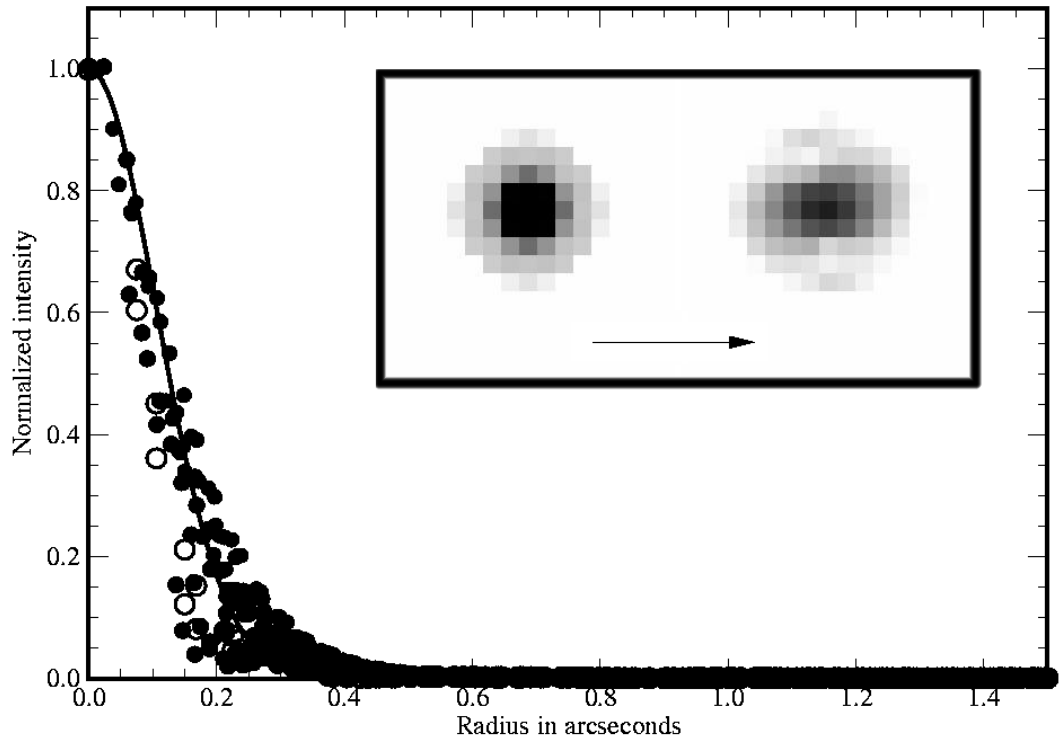}
\caption{Radial profiles of the high-order kernels at 8{\arcsec} offset. The symbols are the same as
Figure~\ref{figure_kernel_tiptilt_20}. Here the arrow indicates the direction to the laser spot.}
\label{figure_kernels_laser_8}
\end{figure}

Even though the high-order Gaussian kernel method requires a model PSF, one might still think of it as a
semiempirical 
case because it is based on a fit to Strehl ratios in the image. This would not be the case with the theoretical
kernel, for example, which would
require a choice for the value $N$ in order to predict the on-axis PSF. Figure~\ref{figure_kernels_laser_8} shows the
average Gaussian high-order
kernel at a radius of 8{\arcsec} from the LGS. For comparison, the theoretical kernel with $N=9$ is also plotted. The
Gaussian and theoretical
kernels appear to be very similar. This suggests that a logical next step might be to predict both the on-axis PSF and
the PSF spatial variation
analytically.

\subsubsection{Semianalytical Methods}

Finally, we employed two semianalytical methods. These approaches are similar to the high-order Gaussian kernel case,
but here we have 
modeled the on-axis PSF as if there were no tip-tilt guide-star offset.
In the case of the Gaussian kernel, for
each frame we estimated the on-axis Strehl ratio from equation 3 assuming an overall throughput ($q=0.5$) and an 
estimated LGS brightness ($V=9$), and then
applying the observed Fried parameter and windspeed. The theoretical kernel used the measurement of the Fried
parameter and $\theta_N$, and assumed a value of $N$. We were guided by the computer simulations to use $N=9$.
We then degraded this PSF due to both
tip-tilt and higher-order anisoplanatism with either the Gaussian or theoretical kernels.  

We label these semianalytical methods, as
opposed to semiempirical ones, because here the on-axis Strehl ratio is predicted with a theoretical model and WFS
measurements, not just
from information in the image itself. In fact, if an independent measurement of $C_n^2$ were available no information
at all would have been
required from the image, and we might then call the method a purely analytical one. No current AO telescope routinely
provides an 
independent $C_n^2$ profile, but our Lick data should at least show if relying on WFS telemetry and a single-layer
$C_n^2$ model can in principle
provide useful results.

\subsection{Predicting the Off-Axis PSF}\label{predicting_the_offaxis_psf}

The fake images were measured in exactly the same manner described in Section~\ref{observations} and compared with the
observations.
Table~\ref{table_psf_predictions} lists the mean percent error in the predicted average Strehl ratio $S$, FWHM, and
ellipticity $\epsilon$ for the
various models
at the positions of the three bright stars. In some cases these predictions are only relevant at the locations of two
stars, one each in the two
arms of the mosaic. In those cases, to simplify the analysis we present the average of the two results. In all cases
we give the absolute percent error, that is, we assume either an overprediction or underprediction of $S$, FWHM, or
$\epsilon$ is equally
undesirable
and have thus dropped the sign. We will deal with the LGS-mode results first, and then compare them with the NGS
results at the end. 

The corrective offset is an improvement in observational technique, albeit a modest one.
The error in Strehl ratio committed by this method was 27{\%} (i.e. $\Delta{S}/S$). The error in FWHM was 26{\%} and the error in
ellipticity 2{\%}. 

The semiempirical approaches improve slightly on the empirical result.
Applying a tip-tilt Gaussian kernel reduces the errors in Strehl ratio and FWHM by a small
amount: to 20{\%} and 12{\%} respectively, with no change in ellipticity.
Similar results were achieved using the theoretical kernel.
The results of the high-order Gaussian method are similar. The mean error in the predicted Strehl ratio is 20{\%} and
FWHM 16{\%}. Note that we did not incorporate anisotropy into the Gaussian kernel model which is why it has done a
poor job of predicting
ellipticity.

Somewhat better results were achieved, at least for predictions of Strehl ratio, by a semianalytical approach.
The Gaussian kernels improved the predicted off-axis Strehl ratio to an accuracy of 14{\%}, and
the theoretical kernels further reduced this to 13{\%}. For a tip-tilt guide star offset as small as 8{\arcsec}, the
results are 
even better; just 2{\%}.
We also applied the same methods while assuming that the height of turbulence for every
observation was the same over both nights ($\bar{h}=5$ km). This did not significantly affect the results,
only increasing the error in
off-axis Strehl ratio at large offsets to 16{\%} (Gaussian) and 15{\%} (theoretical). This suggests that obtaining a
measurement of $\bar{h}$ for every
observation, although helpful, is not necessary.
Useful improvement in PSF characterization may be achieved by obtaining perhaps just one star field mosaic during a
night.

\begin{figure}
\plotone{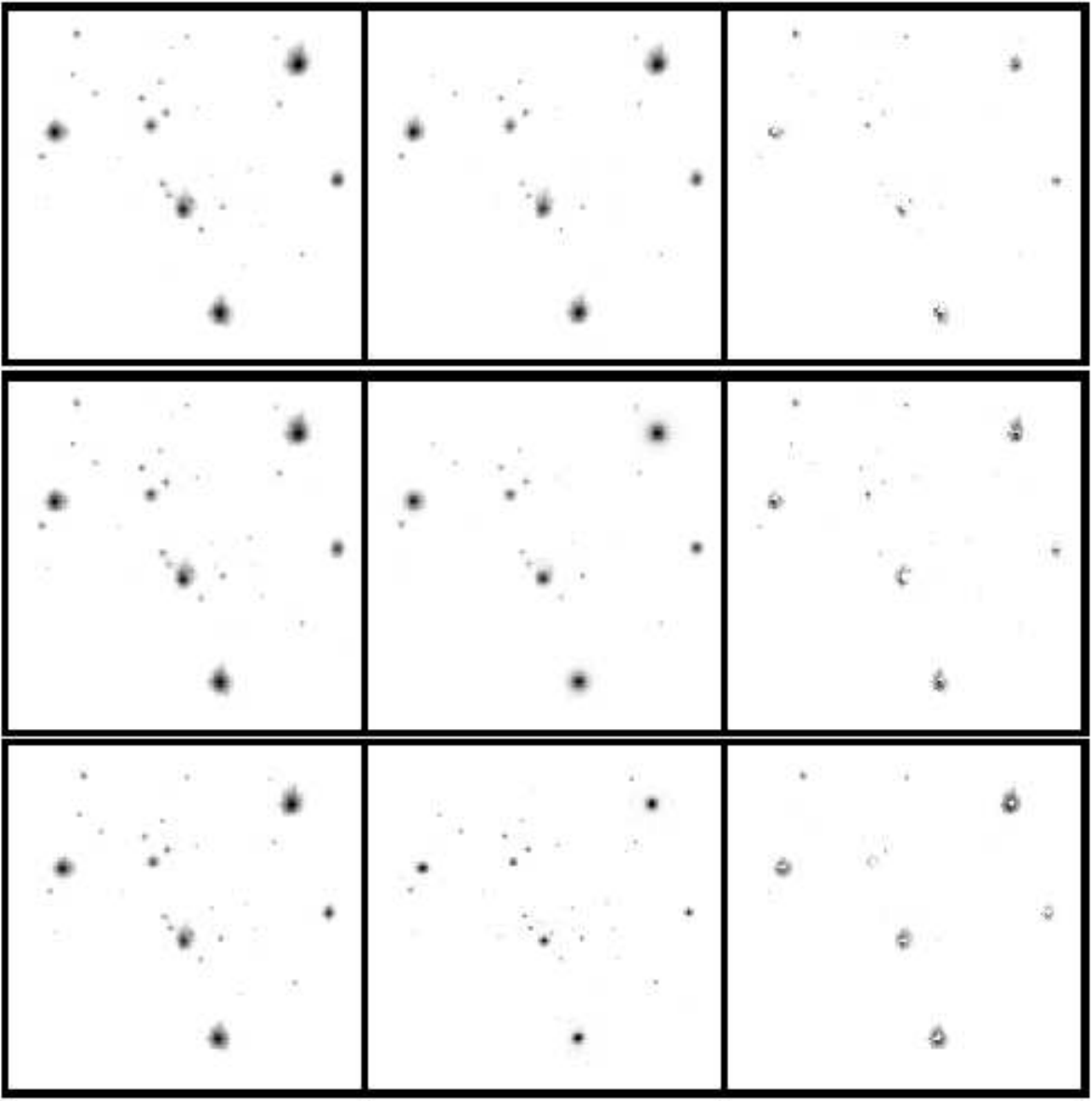}
\caption{The results of three models: empirical corrective offset method (top row) and semianalytical methods with
either Gaussian kernels
(middle) or theoretical kernels (bottom).
Images of the data are along the left column, simulations in the center, and residuals along the right. These are the
results of 
averaging all of the individual images. All are plotted with the same log scale. Note in particular the four brightest
stars, and notice 
the residual halo that remains after subtraction of the semianalytical models. This is well subtracted in the
empirical case, but the semianalytic
models fit the core more accurately
and thus are still better overall.}
\label{figure_results}
\end{figure}

It should be noted that neither of the semianalytical models benefit from an
empirical image of the PSF. This empirical information would be the only way to incorporate non-common path errors or
camera aberrations into the
PSF. Without it the halo is likely to be underpredicted, which is probably why these methods do a better job than the
empirical
ones at predicting Strehl ratio, but do no better in predicting FWHM. This effect can be seen in
Figure~\ref{figure_results}. Here we 
show the data, an empirical model which uses an image of the tip-tilt guide star without any corrective kernel, the
semianalytical
Gaussian model and semianalytical theoretical model, and their respective residuals. Note the diffuse haloes remaining
in the residuals of the
semianalytical models. These are less apparent in the purely empirical treatment, although in that case variation in
the core has not been
accounted for, which is why the semianalytical results are better overall.

Where they were applicable we tried all of the new techniques on our Lick NGS data, and redid the analysis of Paper I.
Note that the percent error in FWHM which we quote in Table~\ref{table_psf_predictions} (14{\%}) for the original
semiempirical technique
is somewhat better than that which we gave in Paper I (19{\%}). This is 
because in our reanalysis we
used only the 5 NGS mosaics with the highest on-axis Strehl ratios to better compare to the higher Strehl ratio LGS
data.
For large offsets ($\sim25${\arcsec}) we obtained similar, if somewhat better results with all the new methods.

Note that all of our data include a corrective offset, which made the analysis more straightforward.
An AO observer need not apply this offset. Conceivably, the only impact would be the addition of an extra step to the
first two semiempirical approaches. After applying the
tip-tilt kernel to the NGS PSF, a high-order kernel would also be applied to account for the LGS AO field
variation.
The other methods would be unchanged.

\section{Conclusions}\label{conclusions}

We have discussed observations of a dense star field for the purpose of characterizing the variation in the LGS AO
PSF at Lick Observatory. The results of our simulations confirm the expectation that the field dependence of the PSF
is complicated, and that the effect of tip-tilt anisoplanatism cannot be ignored. However, a simple atmospheric model
with a thin
layer of turbulence is still a useful model for our data. A dominant layer at $\sim$5 km above the telescope
reproduces
the anisoplanicity seen in the data.

One simple empirical way to characterize LGS data in the absence of a convenient PSF star would be to use the PSF of
the
NGS obtained with the laser pointed directly at the tip-tilt star. This might match the wings well but would 
underestimate the width of the core badly for offsets of 8{\arcsec} or more. According to
our simulations, this would have introduced an error greater than 67{\%} in Strehl ratio (i.e. $\Delta{S}/S$)  for our data.  
A potential improvement in observational technique would be to offset the
tip-tilt guide star by the same amount (and in the opposite direction) as the scientific observations. This
somewhat improves the accuracy in the PSF
prediction, reducing the error in Strehl ratio in our data 27{\%}.

We advocate that LGS AO observers obtain at least one star-field mosaic each night using the same offsets as their
target
observations, and repeat this with high-order correction turned off.
Just as in NGS-mode observations, the mosaicked images must be completed quickly before variations in delivered Strehl
ratio ruin
the measurements.
By using just the information in the tip-tilt-only mosaic and applying a corrective kernel to the empirical PSFs at
the tip-tilt guide star
position, Strehl ratio uncertainty was reduced to 20{\%} (tip-tilt Gaussian kernel) and 18{\%} (tip-tilt theoretical
kernel) for our data.
Another semiempirical method of comparable accuracy (Strehl ratio to within 20{\%}) is possible by instead using information
from the high-order LGS AO mosaic.
All of these semiempirical techniques inherently account for cone effect and should therefore be applicable for AO on
10-meter apertures.

Results with the semianalytical methods were more comparable to NGS AO.
Errors in prediction of either Strehl ratio were better than 14{\%}, and even as good as a few percent if the offset
is small. 
However, their application is complicated by the intertwined effects of $r_0$, $\bar{h}$, $\bar{v}$, $q$, and $N$.
Better results could be realized by separating these effects through independent measurement of $C_n^2$. We do not
find
that the significant computational effort in applying the theoretical kernels was worthwhile for our data because the
prediction of FWHM with the Gaussian kernels is equally good.

\acknowledgements

We would like to thank the staff of Lick Observatory for making these observations
possible. We thank summer student Q. Konopacky for her assistance with the observing log. We also thank T. Fusco and
G. Molodij for their help and
original notes concerning the Zernike polynomial autocorrelations. This work was supported by the
National Science Foundation Science and Technology Center for Adaptive Optics, managed
by the University of California at Santa Cruz under cooperative agreement No.
AST-9876783.

\clearpage

\begin{deluxetable}{lcccccccccccc}
\tablecolumns{13}
\tablecaption{Journal of Observations\label{table_journal_of_observations}}
\tablewidth{0pt}
\tabletypesize{\small}
\rotate
\tablehead{& & & & &\multicolumn{7}{c}{AO correction}\\
\cline{6-13}
& & & & &\multicolumn{5}{c}{Full} & &\multicolumn{2}{c}{Tip-tilt}\\
\cline{6-10} \cline{12-13}
& & & & &\multicolumn{2}{c}{8 arcsec tip-tilt offset} & &\multicolumn{2}{c}{20 arcsec tip-tilt offset} & & &\\
\cline{6-7} \cline{9-10}
&\multicolumn{2}{c}{UT (2001)} & & & &\colhead{$\theta_N$} & & &\colhead{$\theta_N$} & & &\colhead{$\theta_2$}\\
\cline{2-3}
\colhead{Obs.} &\colhead{Date} &\colhead{Time} &\colhead{Air.} &\colhead{$r_0(0.5\mu{\rm m}, 0^o)$} &\colhead{$S$
(on-axis)} &\colhead{(arcsec)} &
&\colhead{$S$ (on-axis)} &\colhead{(arcsec)} & &\colhead{$S$ (on-axis)} &\colhead{(arcsec)}}
\startdata
1 &Aug 16 &08:19 &1.11 &18.6 &$0.10\pm0.04$ &$11\pm4${\phn} & &\nodata &\nodata & &\nodata &\nodata\\ 
2 & &08:27 &1.11 &19.3 &$0.12\pm0.04$ &$14\pm5${\phn} & &$0.14\pm0.03$ &$10\pm3$ & &\nodata &\nodata\\ 
3 & &08:42 &1.12 &19.5 &$0.18\pm0.09$ &$13\pm6${\phn} & &$0.18\pm0.05$ &$12\pm4$ & &\nodata &\nodata\\ 
4 & &08:51 &1.13 &18.6 &\nodata &\nodata & &\nodata &\nodata & &$0.07\pm0.02$ &$40\pm9$\phn\\ 
5 & &09:04 &1.14 &17.2 &$0.29\pm0.03$ &\phn$8\pm1${\phn} & &$0.26\pm0.07$ &$12\pm4$ & &\nodata &\nodata\\ 
6 & &09:21 &1.16 &18.0 &$0.23\pm0.05$ &$10\pm2${\phn} & &$0.26\pm0.15$ &$10\pm2$ & &\nodata &\nodata\\ 
7 & &09:30 &1.19 &19.0 &\nodata &\nodata & &\nodata &\nodata & &$0.09\pm0.02$ &$25\pm6$\phn\\ 
8 & &09:45 &1.22 &21.2 &$0.23\pm0.09$ &$13\pm5$\phn & &$0.33\pm0.09$ &$12\pm5$ & &\nodata &\nodata\\ 
9 & &09:57 &1.25 &22.2 &$0.26\pm0.17$ &$16\pm10$ & &$0.31\pm0.09$ &$16\pm7$ & &\nodata &\nodata\\ 
10 & &10:09 &1.27 &21.8 &$0.44\pm0.19$ &$13\pm6$\phn & &$0.24\pm0.10$ &$12\pm3$ & &\nodata &\nodata\\ 
11 & &10:24 &1.33 &21.1 &\nodata &\nodata & &\nodata &\nodata & &$0.10\pm0.01$ &$44\pm10$\\ 
12 &Aug 17 &07:50 &1.10 &13.3 &$0.11\pm0.05$ &$16\pm8$\phn & &$0.05\pm0.02$ &$10\pm6$ & &\nodata &\nodata\\ 
13 & &08:10 &1.11 &12.0 &$0.09\pm0.02$ &$11\pm2$\phn & &$0.15\pm0.04$ &\phn$9\pm3$ & &\nodata &\nodata\\ 
14 & &08:20 &1.11 &13.1 &$0.11\pm0.05$ &$12\pm5$\phn & &$0.16\pm0.04$ &$11\pm4$ & &\nodata &\nodata\\ 
15 & &08:27 &1.11 &14.3 &\nodata &\nodata & &\nodata &\nodata & &$0.10\pm0.02$ &$26\pm6$\phn\\ 
16 & &08:34 &1.12 &14.3 &$0.08\pm0.01$ &$13\pm2$\phn & &$0.13\pm0.02$ &$10\pm2$ & &\nodata &\nodata\\ 
17 & &08:50 &1.13 &11.4 &$0.08\pm0.01$ &$14\pm1$\phn & &$0.09\pm0.02$ &$17\pm7$ & &\nodata &\nodata\\ 
18 & &09:02 &1.15 &~9.8 &$0.10\pm0.02$ &$11\pm3$\phn & &$0.13\pm0.03$ &\phn$9\pm3$ & &\nodata &\nodata\\ 
19 & &09:09 &1.16 &~9.8 &\nodata &\nodata & &\nodata &\nodata & &$0.05\pm0.01$ &$30\pm6$\phn\\
20 & &09:25 &1.18 &10.1 &\nodata &\nodata & &$0.11\pm0.01$ &\phn$9\pm1$ & &\nodata &\nodata\\ 
21 & &09:33 &1.20 &10.1 &\nodata &\nodata & &$0.14\pm0.05$ &$11\pm6$ & &\nodata &\nodata\\ 
22 & &09:42 &1.22 &10.4 &$0.12\pm0.09$ &$24\pm18$ & &$0.12\pm0.01$ &$12\pm5$ & &\nodata &\nodata\\ 
23 & &09:49 &1.24 &10.9 &\nodata &\nodata & &\nodata &\nodata & &$0.07\pm0.01$ &$25\pm6$\phn\\ 
24 & &10:01 &1.27 &11.9 &$0.24\pm0.07$ &$14\pm4$\phn & &$0.15\pm0.06$ &$17\pm7$ & &\nodata &\nodata\\
25 & &10:11 &1.30 &14.7 &$0.10\pm0.07$ &$29\pm19$ & &$0.14\pm0.03$ &$12\pm3$ & &\nodata &\nodata\\ 
\enddata
\end{deluxetable}

\clearpage

\begin{deluxetable}{lcccccccc}
\tablecaption{Comparison to Computer Modeling Results ($\bar{h}=5~{\rm km}$, $N=9$)\label{table_computer_modeling}}
\tablewidth{0pt}
\tabletypesize{\small}
\tablehead{& & & &\multicolumn{5}{c}{$r_0(0.5~\mu{\rm m})~{\rm cm}$}\\
\cline{5-9}
& & & &\multicolumn{2}{c}{20} & &\multicolumn{2}{c}{12}\\
\cline{5-6} \cline{8-9}
&\multicolumn{2}{c}{Observed values} & &\multicolumn{2}{c}{$q$} & &\multicolumn{2}{c}{$q$}\\
\cline{2-3} \cline{5-6} \cline{8-9}
\colhead{Quantity} &\colhead{August 16} &\colhead{August 17} & &\colhead{0.33} &\colhead{0.67} & &\colhead{0.33}
&\colhead{0.67}}
\startdata
\sidehead{Mean of 8 and 20 arcsec tip-tilt offsets (2.2~$\mu{\rm m}$)}
$S_0$ &$0.24\pm0.09$\tablenotemark{1} &$0.12\pm0.04$\tablenotemark{1} & &0.28 &0.40 & &0.12 &0.18\\
$\theta_N$ (arcsec) &$12\pm2$\tablenotemark{1} &$14\pm4$\tablenotemark{1} & &16 &14 & &12 &9\\
\sidehead{No tip-tilt offset (2.2~$\mu{\rm m}$)}
$S_0$ &\nodata &\nodata & &0.32 &0.41 & &0.13 &0.22\\
$\theta_N$ (arcsec) &\nodata &\nodata & &13 &12 & &9 &7\\
\enddata
\tablenotetext{1}{Standard deviation of the mean.}
\end{deluxetable}

\clearpage

\begin{deluxetable}{llccc}
\tablecaption{Summary of Methods\label{table_model_summary}}
\tablewidth{0pt}
\tabletypesize{\small}
\rotate
\tablehead{\colhead{} & &\multicolumn{3}{c}{In LGS mode, accounts for:}\\
\cline{3-5}
\colhead{Name} & &\colhead{Cone} &\multicolumn{2}{c}{Anisoplanatism}\\
\cline{4-5}
\colhead{(Correction applied)} &\colhead{Description} &\colhead{effect} &\colhead{Tip-tilt} &\colhead{High-order}}
\startdata
Empirical & & & &\\
~~~No correction &Uses image of the NGS to predict off-axis PSF; not tested here &yes &no &no \\
~~~Corrective offset &Same as above, but with NGS shifted to correct for target offset &yes &no &yes\\
Semiempirical & & & &\\
~~~Deconvolved kernel &Kernel deconvolved from mosaic; applies only to NGS mode &\nodata &\nodata &\nodata\\
~~~Tip-tilt Gaussian kernel &Improves on corrective offset by applying a tip-tilt kernel &yes &yes &yes\\
~~~Tip-tilt theoretical kernel &Same as above, but using a theoretical kernel &yes &yes &yes\\
~~~High-order Gaussian kernel &On-axis PSF given by fit to mosaic; off-axis kernel is Gaussian &yes &yes &yes\\
Semianalytical & & & &\\
~~~Both Gaussian kernels &Predicts NGS PSF from telemetry; kernels are Gaussian &no\tablenotemark{1} &yes &yes\\
~~~Both theoretical kernels &Same as above, but using theoretical kernels &no &yes &yes\\
\enddata
\tablenotetext{1}{Cone effect could be incorporated here by including it in equation 3.  It is a small effect for our
data.}
\end{deluxetable}

\clearpage

\begin{deluxetable}{lccccccccccc}
\tablecaption{Mean Percent Errors in Off-Axis PSF Predictions\label{table_psf_predictions}}
\tablewidth{0pt}
\tabletypesize{\small}
\rotate
\tablehead{\colhead{} &\multicolumn{3}{c}{NGS mode} & &\multicolumn{7}{c}{LGS mode}\\
\cline{6-12}
\colhead{} &\multicolumn{3}{c}{25 arcsec offset} & &\multicolumn{3}{c}{$20+8$ arcsec offset} &
&\multicolumn{3}{c}{$8+8$ arcsec offset}\\
\cline{2-4} \cline{6-8} \cline{10-12}
\colhead{Method} &\colhead{$S$} &\colhead{FWHM} &\colhead{$\epsilon$} & &\colhead{$S$} &\colhead{FWHM}
&\colhead{$\epsilon$} & &\colhead{$S$}
&\colhead{FWHM} &\colhead{$\epsilon$}}
\startdata
Empirical & & & & & & & & & & &\\
~~~No correction &42 &29 &~2 & &\nodata &\nodata &\nodata & &\nodata &\nodata &\nodata\\
~~~Corrective offset &\nodata &\nodata &\nodata & &27 &26 &~2 & &\nodata &\nodata &\nodata\\
Semiempirical & & & & & & & & & & &\\
~~~Deconvolved kernel &14 &14 &~2 & &\nodata &\nodata &\nodata & &\nodata &\nodata &\nodata\\
~~~Tip-tilt Gaussian kernel &13 &~5 &~1 & &20 &12 &~2 & &\nodata &\nodata &\nodata\\
~~~Tip-tilt theoretical kernel &10 &~4 &~2 & &18 &16 &~2\ & &\nodata &\nodata &\nodata\\
~~~High-order Gaussian kernel &\nodata &\nodata &\nodata & &20 &16 &13 & &~4 &~3 &16\\
Semianalytical & & & & & & & & & & &\\
~~~Both Gaussian kernels &~9 &14 &10 & &14 &16 &13 & &~5 &~1 &16\\
~~~Both theoretical kernels &~8 &~4 &~7 & &13 &18 &~6 & &~2 &~2 &~1\\
\enddata
\end{deluxetable}

\clearpage

\clearpage

\clearpage

\clearpage

\clearpage

\end{document}